\documentclass[12pt]{article}
\usepackage{graphicx}
\usepackage[utf8]{inputenc}  
\usepackage[T1]{fontenc}
\usepackage{amsfonts,amssymb,amsthm,amsmath}
\usepackage[french,english]{babel}
\usepackage{graphics,graphicx}

\begin{document}

\title{Special Relativity. Using tangent boost along a worldline and its associated matrix in the Lie algebra of the Lorentz group. Applications}
\author{M. Langlois\\
Passavant, 25360, France\\ 
{\em mj.langlois@wanadoo.fr}\\\\
M. Meyer\\
Laboratoire de Mathématiques\\
Université de
Franche-Comté, 25030 Besançon Cedex, France\\
{\em mmeyer@univ-fcomte.fr}\\\\
J.-M. Vigoureux\\
Institut UTINAM, UMR CNRS 6213,\\Université de
Franche-Comté, 25030 Besançon Cedex, France.\\
{\em jean-marie.vigoureux@univ-fcomte.fr}}
\maketitle 

\begin{abstract}
In order to generalize the relativistic notion of boost to the case of non inertial particles and to general relativity, we look closer into the definition of the Lie group of Lorentz matrices and its Lie algebra and we study how this group acts on the Minskowski space. We thus define the notion of tangent boost along a worldline. This very general notion gives a useful tool both in special relativity (for non inertial particles or/and for non rectilinear coordinates) and in general relativity. We also introduce a matrix of the Lie algebra which, together with the tangent boost, gives the whole dynamical description of the considered system (acceleration and Thomas rotation). After studying the properties of Lie algebra matrices and their reduced forms, we show that the Lie group of special Lorentz matrices has four one-parameter subgroups. These tools lead us to introduce the Thomas rotation in a quite general way. At the end of the paper, we present some examples using these tools and we consider the case of an electron rotating on a circular orbit around an atom nucleus. We then discuss the twin paradox and we show that when the one who made a journey into space in a high-speed rocket returns home he is not only younger than the twin who stayed on Earth but he is also disorientated because his gyroscope has turned with respect to earth referential frame.
\end{abstract}

\section{Introduction}
In the frame of special relativity theory the history of an inertial particle is described by a geodesic straight line in the four dimensional Minkowski space, endowed with the $\eta = diag (-1, 1, 1, 1)$ metric. This geodesic is a timelike straight line and its orthogonal complement is the physical space of the particle formed by all its simultaneous events.
The passage from one inertial particle to another one is done through a special Lorentz matrix, which is called the boost, and this process is the Lorentz-Poincaré transformation.
But for a non inertial particle, all this is lost since its worldline is no more a straight line and there is no Lorentz transformation and boost associated to it. In order to fill this gap we suggest a deeper insight into the action of the Lie group of Lorentz matrices (and its Lie algebra) on the Minkowski space. This leads us to a new definition of a\textit{ tangent boost} along a worldline. This notion may be used in both situations of special or general relativity theories. Therefore we introduce a matrix belonging to the Lie algebra, which, together with the tangent boost, describes completely the dynamical system: acceleration and instantaneous Thomas rotation.\\
In a first part, we present properties of Lie matrices and of their reduced forms and we show that the Lie group of special and orthochronous Lorentz matrices has four one-parameter subgroups. These tools permit to introduce the Thomas rotation in a quite general way. Then, we give some applications of these tools: we first consider the case of an uniformly accelerated system and the one of an electron rotating on a circular orbit around the atom nucleus. We then present the case of the so-called "Langevin's twins" and we show that, when the twin who made a journey into space returns home, he is not only younger than the twin who stayed on Earth but he is also disorientated with respect to the terrestial frame because his gyroscope has turned with respect to the earth referential frame.\\
Let us underline that this formalism can be used both in Special and in General Relativity

\section{The Lie algebra of a Lie group}
A Lie group is a smooth manifold with a compatible group structure, which means that the product and inverse operations are smooth. The Lie algebra of this Lie group can be seen as the tangent space $T_e$ to the manifold at the unit element \textit{e} of the group multiplication. This tangent space is a vector space endowed with the Lie bracket of two tangent vectors.
 \subsection{Example : The Lie group $SO(3)$ and its Lie algebra}
 Let's start with the group of $3\!\times\! 3$-matrices having $+1$ determinant. As a smooth manifold, it can be regarded as a 3-dimensional submanifold of 
 $\mathbb{R}^9$ defined by the 6 equations resulting from the orthogonal matrix definition : $^TAA=I$. Let's denote it, as usual, by $SO(3)$.\\
 Its Lie algebra $\mathcal{A}$ is the 3-dimensional vector space of skew-symmetric matrices endowed with the bracket
 $$[\Omega_1,\Omega_2] = \Omega_1.\Omega_2 -\Omega_2.\Omega_1$$ 
 This manifold is obviously isomorphic to the euclidean space $\mathbb{R}^3$ endowed with the cross product. The vectors of our Lie algebra should be regarded as tangent vectors $\displaystyle{\frac{dA}{dt} }\in T_A$ of smooth paths $t \rightarrow A(t)$ on the $SO(3)$ manifold. The left translation on the group by $A^{-1}=  {^T\!\!A}$ shall be $L_{A^{-1}} B = {^T\!\!A} B$.
Of course if $A=B$, $L_{A^{-1}}(A) = {^T\!\!A} A= I$. The linear mapping $L_{A^{-1}}$ which by definition is equal to its differential $D L_{A^{-1}}$ maps the tangent vector $\displaystyle{\frac{dA}{dt}}$ from the tangent space $T_A$ to the tangent space $T_I$ :
\begin{equation}\label{un}
\Omega= D L_{A^{-1}}\frac{dA}{dt}\,= A^{-1} \frac{dA}{dt} =\,^T\!\!A \,\frac{dA}{dt} 
\end{equation}
Derivating the relation ${^T\!\!A}A=I$ gives $^T\Omega + \Omega= 0$, which means that $\Omega$ is skew-symmetric. Of course it would be possible to obtain the same $T_I$ through right translation by $A^{-1}$, using
 $R_{A^{-1}} B = B \,{^T\!\!A}$.

\subsubsection*{Application to kinematics of rotation of a rigid body around a fixed point.}
Keeping in mind later comparisons, we shall apply the results of the previous section to the study of the motion of a rigid body.
We want to show the interest of looking at the action of $SO(3)$ on $\mathbb{R}^3$ when this group is regarded as a subgroup of isometries of the euclidean space $\mathbb{R}^3$, and the meaning of its Lie algebra $\mathcal{A}=T_I$. 
Let us associate to a three-dimensional point $O$ two coordinate systems $(O, e)$ and $(O, E)$ defined by two orthonormal basis $e$ and $E$ respectively. And let $A$ denote the matrix mapping $e$ to $E$. A smooth path $t \rightarrow A(t)$ on the $SO(3)$ manifold corresponds to a rotation movement of $(O, E)$ around $O$ with respect to the coordinate system $(O, e)$.\\
Let us denote by $X_e$ and $X_E$ the coordinates of a point $M$ in the neighborhood of $O$ in the reference systems $(O, e)$ and $(O, E)$ respectively. Looking at the movement of $M$ with respect to $(O, E)$, we then have 
$$X_e(t) = A(t)  X_E(t) \quad\Rightarrow \quad \frac{dX_e}{dt} = A\, \frac{dX_E}{dt}+\frac{dA}{dt} \, X_E$$

composing by the left translation $A^{-1}$ and using (\ref{un}) we get
$$A^{-1}\frac{dX_e}{dt} =  \frac{dX_E}{dt}+A^{-1}\,\frac{dA}{dt}  \, X_E =  \frac{dX_E}{dt}+\Omega_E \, X_E$$
This relation expresses the derivation rule of the movement of a point $X$ in the moving coordinate system $(M, E)$. The absolute derivative 
$\displaystyle{\left(\frac{dX}{dt}\right)}_e$ of X with respect to $t$ is equal to the sum of the relative derivative  $\displaystyle{\left(\frac{dX}{dt}\right)}_E$ and of the training derivative defined by $\Omega_E X_E$

\begin{equation}
\label{eq1}
\displaystyle{\left(\frac{dX}{dt}\right)}_e=\displaystyle{\left(\frac{dX}{dt}\right)}_E + \Omega_E X_E
\end{equation}
$\Omega_E$ is a skew-symmetric covariant tensor whose adjoint gives the components of a vector  $\omega$ in $\mathbb{R}^3$ and permits us to express the training velocity in the well known vector form : 
\begin{equation}\label{omegaX}
\Omega_E X_E= (\omega \wedge X)_E
\end{equation} \\
An analogous process starting from the right translation $R_{A^{-1}}$ would lead us to the same derivation rule as given in (\ref{eq1}) but using the components in $(M, e)$. 
Note that 
$$\displaystyle{\frac{dA}{dt}}A^{-1}=\Omega_e$$
There is another interesting application of the identification $\mathcal{A}=T_I \approx \mathbb{R}^3$ : 
The exponential map $$\Omega \rightarrow \exp{\Omega}= \sum \frac{\Omega^k}{k !}$$ is a diffeomorphism of the open ball $\mathcal{B}$ of radius $\pi$ in $\mathbb{R}^3$ into an open subset of $SO(3)$, and we thus obtain an interesting parametrization of $SO(3)$ : writing $\varpi$ the norm of $\omega$  and 
$A=\exp\Omega \in SO(3)$, we have 
$$A X = X + (\sin\varpi )\,\, \omega \wedge X+ (\cos \varpi- 1) \,\omega \wedge(\omega \wedge X)
 \qquad X\in \mathbb{R}^3.$$ 
Such a formula has an obvious geometrical meaning : $\exp\Omega$ is a rotation through the angle $\varpi$ about the axis having the direction of the vector $\omega$.
Note that, $\Omega$ being independant of $t$, the function $t\rightarrow \exp{(t\Omega} ) : \mathbb{R}\rightarrow SO(3)$ defines a one-parameter subgroup of $SO(3)$, and that the matrix product $\mathbf{\exp}{(t\Omega}).X_0 $ is the solution of the linear differential equation $$\frac{dX_e}{dt}=\Omega_e.X_e$$ with the initial condition 
$X_e =X_0$. This equation is nothing but (\ref{eq1}) when written in the coordinate system $(M, e)$. Its solution defines a uniform rotation.\\
\section{The Lie group of Lorentz matrices. \\ Application to special relativity}
\subsection{Preliminaries}
In special relativity the motion of an inertial particle with respect to an inertial observer is described by a Lorentz-Poincaré transformation.
This transformation is associated to a $4\!\times\! 4$-matrix belonging to the subgroup of orthochronous Lorentz matrices of $+1$ determinant : they map the subset of timelike future oriented vectors into itself. They transform a $\eta$-orthonormal basis (associated to the inertial observer) into another $\eta$-orthonormal basis (associated to the particle).\\
The columns of such a matrix have a clear physical and geometrical interpretation : the first column is the \textit{4-velocity} of the particle (a unitary timelike \textit{4-vector} tangent to the worldline), and the three other columns define an orthonormal basis of the physical space of the particle.
We turn now to the more general situation of a non inertial particle : the relative motion between two non inertial particles, or between a non inertial particle and another (inertial or non inertial) observer will be described by a time-dependent function with values in the group of Lorentz transformations. We thus naturally come to the notion of \textit{tangent boost along a worldline}, we shall now study its main properties.

\subsection{The Lie group of Lorentz matrices and its associated Lie algebra}
The shall denote by $\mathcal{S}$ the subgroup of the Lie group of Lorentz matrices consisting of all orthochronous (Lorentz) matrices with $+1$ determinant. 
It is a 6-dimensional submanifold of $\mathbb{R}^{16}$ as defined by the 10 equations involving the 16 coordinates $s_{ij}$ of the matrix $S$, obtained from the relation  $^T S\,\eta\, S = \eta$.\\
This group $\mathcal{S}$ acts as a group of isometries on the Minkowski space $\mathcal{M} = (\mathbb{R}^{4},\eta)$.
We shall now be interested in the tangent space $\mathcal{A}=T_{I}$ of $\mathcal{S}$ taken at the identity matrix $I$.\\

Let $t  \rightarrow S(t)$ be a smooth curve on the manifold $ \mathcal{S}$,  and $\displaystyle{\dot{S}=\frac{dS}{dt}}$  its tangent vector belonging to $T_{S(t)}$, the element  $\Lambda=DL_{S^{-1}} (\dot{S})$ of $\mathcal{A}=T_{I}$ shall simply be : 
$$\Lambda= DL_{S^{-1}}(\dot{S}) =L_{S^{-1}} \dot{S}=S^{-1} \dot{S}\in T_I$$
From the relation ${^T\!}S\, \eta\, S=\eta$ we deduce that  $S^{-1}= \eta\, {^T}\!S\, \eta$ and that the covariant tensor of type $(0,2)$, $\Omega ={^T}\!S\,\eta\,\dot{S}$ is skew-symmetric (this is obtained by derivating the relation ${^T}\!S\,\eta\,S$), and so the element $\Lambda$ of the Lie algebra can be rewritten :
\begin{equation}
\Lambda=S^{-1} \dot{S} = \eta\,\Omega\,  \qquad \text{with} \qquad \Omega = {^T}\!S\,\eta\,\dot{S}
\end{equation}

As a conclusion to this subsection, the Lie algebra $\mathcal{S}$ is the linear space of $\eta\,\Omega$ matrices, where $\Omega$ is a skew-symmetric covariant tensor of type $(0,2)$.\\
The skew-symmetric tensor associated to the Lie bracket $[\Lambda_1, \Lambda_2]$ is $$\Omega_1\,\eta\,\Omega_2-\Omega_2\,\eta\,\Omega_1$$

The exponential mapping from the Lie algebra to the group 
\begin{equation}
\Lambda  \rightarrow S = \exp(\Lambda) = \sum \frac{\Lambda^k}{k !}
\end{equation}
 defines a diffeomorphism from $\mathbb{R}^3 \times \mathcal{B}$ into $\mathbb{R}^6$
(recall $\mathcal{B}$ is the open ball of radius $\pi$ in $\mathbb{R}^3$).

\subsection{Properties of the Lie algebra matrices}
Every matrix belonging to the Lie algebra $\mathcal{A}$ can be written
\begin{equation}\label{lambda}
\Lambda(A,B)=\left(\small
\begin{array}{cccc}
 0 &a_1  & a_2 & a_3 \\
a_1  & 0 & -b_3 & b_2 \\
a_2 & b_3 & 0 & -b_1  \\
a_3 & -b_2 & b_1  & 0
\end{array}
\right)
\end{equation}
where  $A=(a_1,a_2,a_3)$ and $B=(b_1,b_2,b_3)$ are spacelike vectors.\\
Note the relation $$\det{\Lambda(A,\,B)= - (A.B)^2}$$
We then have the following proposition about the reduced forms of the matrices :
\textit{Given any matrix $\Lambda$ of $\mathcal{A}$ such that  $A.B \neq  0$ (usual inner product), there exists an $\eta$-orthonormal basis $E=(E_0,E_1,E_2,E_3)$ with respect to which $\Lambda$ can be written} :
\begin{equation}\label{fr1}
\left(\small
\begin{array}{cccc}
 0 & \alpha  & 0 & 0 \\
 \alpha  & 0 & 0 & 0 \\
 0 & 0 & 0 & -\omega  \\
 0 & 0 & \omega  & 0
\end{array}
\right)
\end{equation}
\textit{where $\alpha$ and $\omega$ are two real numbers and where $E_0$ is a timelike $4$-vector, the three other $4$-vectors $E_i$ are spacelike.}\\
\textit{If $A.B = 0$, setting $A.A=a^2$ and $B.B=b^2$, the three reduced matrix forms, according to the three conditions $a^2- b^2 > 0$ , $ a^2- b^2 < 0$ and $b^2 = a^2$ respectively, shall be}

\begin{equation}\label{fr2}
\left(\small
\begin{array}{cccc}
 0 & a  & 0 & 0 \\
 a  & 0 & 0 & 0 \\
 0 & 0 & 0 & 0 \\
 0 & 0 &0  & 0
\end{array}
\right)\qquad \left(\small
\begin{array}{cccc}
 0 & 0  & 0 & 0 \\
0  & 0 & 0 & 0 \\
 0 & 0 & 0 & -b  \\
 0 & 0 & b  & 0
\end{array}
\right) \qquad \left(\small
\begin{array}{cccc}
 0 & a  & 0 & 0 \\
a  & 0 & 0 & -a \\
 0 & 0 & 0 &0  \\
 0 & a & 0  & 0
\end{array}
\right)
\end{equation}
\textbf{Proof}\\
We shall use following notations :
\begin{description}
  \item[n1)] Let $(e_0,e_1,e_2,e_3)$ be a $\eta$-orthonormal basis consisting of $4$-vectors, where $e_0$ is a timelike vector and where $(e_1,e_2,e_3)$ is the basis of the orthogonal complement of the straight line $\mathbb{R}e_0$.

 \item[n2)]To every $3$-vector $V=(v_1,v_2,v_3)$ a $4$-vector $qV = (0, v_1,v_2,v_3)$ can be associated, which obviously is space like. Note that this process leeds us to the definition of a linear mapping $q$.\\Any $4$-vector $\textbf{V}=(v_0,v_1,v_2,v_3)=(v_0,V)$ can be written $\textbf{V}=v_0\,e_0+qV$.
The inner product of two $4$-vectors $\textbf{U}$  and $\textbf{V}$ shall be written $\langle \textbf{U},\textbf{V}\rangle$, and we have the formula :
$$\langle \textbf{U},\textbf{V}\rangle=-u_0\,v_0+U.V$$ 
 \item[n3)] With the aim of more elegant computations we shall write $C$ the cross product $B\times A$, and $a, b, c$ the euclidean norms of the $3$-vectors $A,B$ and $C$ respectively. The following formulas will be useful : $$c^2=(B\times A)^2=a^2b^2-(A.B)^2,\quad B\times C=(A.B)B-b^2A$$
  \item[n4)] With the aim of studying the action of $\Lambda$ on $4$-vectors, we write $\Lambda(A,B)$ as $\Lambda_A+\Lambda_B$ : 
\begin{equation}\label{lambda2}
\Lambda_A+\Lambda_B=\left(\small
\begin{array}{cccc}
 0 &a_1  & a_2 & a_3 \\
a_1  & 0 & 0 & 0\\
a_2 & 0 & 0 & 0  \\
a_3 & 0 & 0  & 0
\end{array}
\right)+
\left(\small
\begin{array}{cccc}
 0 & 0  & 0 & 0 \\
0  & 0 & -b_3 & b_2 \\
0 & b_3 & 0 & -b_1  \\
0 & -b_2 & b_1  & 0
\end{array}
\right)
\end{equation}

Let $\textbf{V}=(v_0,v_1,v_2,v_3)$ be any $4$-vector, we get the following formulas for the matrix products :
\begin{eqnarray}\label{FV}
\Lambda_A.\mathbf{V} & = & A.Ve_0+v_0\,qA\nonumber \\
\Lambda_B.\mathbf{V} & = & q(B\times V)\\
\Lambda .\mathbf{V}& = & A.Ve_0+q(v_0A+B\times V)\nonumber 
\end{eqnarray} 
\end{description}

To obtain the reduced form of $\Lambda$ lets start with the study of $\Lambda^2$. Indeed its characteristic polynomial simply factorizes 
$P(X) =(P_m(X))^2$
where
\begin{equation}\label{polmin}
Pm(X)  =  X^2-(A^2-B^2)X-(A.B)^2 
\end{equation}
$P_m(X)$ is the minimal polynomial of the matrix $\Lambda^2$, which means that we have the matrix relation $P_m(\Lambda^2)=O$
\begin{eqnarray}\label{pc}
 P_m(X)& = & (X-\alpha^2)(X+\omega^2) \nonumber\\
 P_m(\Lambda^2)&=&(\Lambda^2-\alpha^2 I)(\Lambda^2+\omega^2 I)=O
\end{eqnarray}
where $\alpha^2$ and $-\omega^2$ are the two zeros of $P_m(X)$.
We wrote $I$ and $O$ for the identity and the zero $4\times4$-matrices.\\
Note by the way the formulas linking the roots of the polynomial (\ref{polmin})):
\begin{eqnarray}\label{rel1}
A^2-B^2 & = & a^2-b^2 = \alpha^2-\omega^2\nonumber\\
A.B & = & \alpha\,\omega
\end{eqnarray}
The first columns of the matrices $\Lambda^2-\alpha^2 I$ and $\Lambda^2+\omega^2I$ 
are the 4-vectors obtained by computing the product $\Lambda .qA$ respectively, using (\ref{FV}) :
\begin{eqnarray}\label{col1}
\Lambda .qA -\alpha^2e_0 & = & -\alpha^2e_0 +(A.A)\,e_0+B\times A= (a^2-\alpha^2)\,e_0+C\nonumber\\
\Lambda .qA +\omega^2e_0& = & (a^2+\omega^2)\,e_0+C
\end{eqnarray}
The relation (\ref{pc}) means that the columns of the matrix $(\Lambda^2+\omega^2 I)$ generate the eigenspace of $\Lambda^2$ associated to the eigenvalue $\alpha^2$ and that the columns of the matrix $(\Lambda^2-\alpha^2 I)$ generate the eigenspace associated to $-\omega^2$. 
Let us write $\Pi_\alpha$ and $\Pi_\omega$ these two 2-dimensional eigenspaces.
\begin{description}
\item[ The eigenspace $\Pi_\alpha $ associated to the eigenvalue $\alpha^2$]:\\
Writing $\mathbf{W}_1 \in \Pi_\alpha$ the vector defined by the first column of $(\Lambda^2+\omega^2 I)$ and $\mathbf{W}_2 = \Lambda .\mathbf{W}_1$, $(\mathbf{W}_1,\mathbf{W}_2)$  is an $\eta$-orthogonal basis of $\Pi_\alpha$. 
Indeed, on the one hand :
\begin{equation}
\label{ rel2}
\Lambda .\mathbf{W}_2 = \Lambda^2.\mathbf{W}_1=\alpha^2 \mathbf{W}_1 \quad \Rightarrow \quad  \Lambda^2 .\mathbf{W}_2=\alpha^2\Lambda .\mathbf{W}_1= \alpha^2 \mathbf{W}_2
\end{equation}
shows that $\mathbf{W}_2$ belongs to $\Pi_\alpha$ and on the other hand, using (\textbf{n3}) to compute $\textbf{W}_2$ we get :
\begin{eqnarray}\label{W}
\mathbf{W}_1 & = & (a^2+\omega^2)\, e_0+qC =b^2 e_0+qC\nonumber\\
\mathbf{W}_2&=&\Lambda .\mathbf{W}_1=A.C e_0+q((a^2+\omega^2)A+B\times C)\\
&=&q((\omega^2+a^2)A-b^2A +(A.B)B)=\alpha q(\alpha A+\omega B)\nonumber
\end{eqnarray}
Apart from the relation $\langle \mathbf{W}_1,\mathbf{W}_2\rangle=0$ we also have  :
\begin{eqnarray}\label{NW}
\langle \mathbf{W}_1,\mathbf{W}_1\rangle & = & -(\alpha^2+\omega^2)(a^2+\omega^2)=-N_1^2\nonumber \\
\langle \mathbf{W}_2,\mathbf{W}_2\rangle & = & \alpha^2(\alpha^2+\omega^2)(a^2+\omega^2)=N_2^2
\end{eqnarray}
which means that $\mathbf{W}_1$ is timelike and that $\mathbf{W}_2$ is spacelike. Also note the relation $N_2 = \alpha N_1$.

Writing now $E_0=\frac{1}{N_1}\mathbf{W}_1$, $E_1=\frac{1}{N_2}\mathbf{W}_2$, $(E_0,E_1)$ defines an orthonormal basis of the space like plane $\Pi_\alpha$ with:
\begin{eqnarray}\label{rel3}
\Lambda .E_0 & = & \frac{1}{N_1} \Lambda .\mathbf{W}_1=\frac{1}{N_1} \mathbf{W}_2=\frac{N_2}{N_1}E_1=\alpha E_1\nonumber\\
\Lambda .E_1 & = & \frac{1}{N_2} \Lambda .\mathbf{W}_2=\frac{1}{N_2} \Lambda^2 .\mathbf{W}_1=\alpha^2\frac{N_1}{N_2} \mathbf{W}_1=\alpha E_0
\end{eqnarray}
 \item[The eigenspace $\Pi_\omega $ associated to the eigenvalue $-\omega^2$]:\\
Writing $\mathbf{V}_1 \in \Pi_\omega$ the vector defined by the first column of $(\Lambda^2-\alpha^2 I)$ and $\mathbf{V}_2 = \Lambda .\mathbf{V}_1$, $(\mathbf{V}_1,\mathbf{V}_2)$  is an $\eta$-orthogonal basis of $\Pi_\omega$. 
Moreover $\mathbf{V}_ 1$ and $\mathbf{V}_2$ are spacelike and $\Pi_\omega$ is the orthogonal complement of $\Pi_\alpha$. 
Here is an outline of the computations:
\begin{eqnarray*}
\mathbf{V}1 & = & (a^2-\alpha^2) e_0+qC\\
\mathbf{V}2 & = &\Lambda .\mathbf{V}_1=q((a^2-b^2-\alpha^2)A+\alpha\,\omega B)\\
& = &\omega \,q(-\omega A+\alpha B)\nonumber\\
\Lambda .\mathbf{V}_2 &=& \Lambda^2 .V_1=-\omega^2 \mathbf{V}_1 \quad\Rightarrow \quad  \Lambda^2 .\mathbf{V}_2=-\omega^2\Lambda .\mathbf{V}_1= -\omega^2 \mathbf{V}_2
\end{eqnarray*}
Apart from the relation $\langle \mathbf{V}_1,\mathbf{V}_2\rangle=0$ we also have:
\begin{eqnarray*}
\langle \mathbf{V}_1,\mathbf{V}_1\rangle & = & (\alpha^2+\omega^2)(a^2-\alpha^2)=N_1^2 \\
\langle \mathbf{V}_2,\mathbf{V}_2\rangle & = &\omega^2(\alpha^2+\omega^2)(a^2-\alpha^2)=N_2^2\\
N_2 & = & \omega N_1
\end{eqnarray*}
The plane $\Pi_\omega$ is obviously spacelike, and writing:
$$E_2=\frac{1}{N_1}\mathbf{V}_1, \quad E_3=\frac{1}{N_2}\mathbf{V}_2$$
$(E_2,E_3)$ constitutes an orthonormal basis of $\Pi_\omega$ with the relations:
\begin{eqnarray}\label{rel4}
\Lambda .E_2 & = & \omega\,E_3\nonumber\\
\Lambda .E_1 & = & -\omega\, E_2
\end{eqnarray}
These two relations (\ref{rel4}), with the former relations  (\ref{rel3}) linking $E_0$ and $E_1$, show that the matrix $\Lambda$ gets the reduced form (\ref{fr1}) in the $\eta$-orthonormal basis $(E_0,E_1,E_2,E_3)$. 
Let us recall that $E_0$ is timelike and that the three other $(E_1,E_2,E_3)$ are spacelike; they define an orthonormal basis of $\mathbb{R}^3$, the orthogonal complement of the line $\mathbb{R}E_0$.
\item[The situation where A and B are orthogonal] :
In the case $(A.B=0)$ we need to discuss according to the sign of $a^2-b^2$ since the two roots of (\ref{pc}) are $\alpha^2=a^2-b^2$ and $\omega^2=0$ and matrices $\Lambda^2$ and $\Lambda$ are of rank 2. The minimal polynomial (\ref{polmin}) can be simplified and the relation (\ref{pc}) becomes: 
$$Pm(\Lambda^2)=(\Lambda^2-\alpha^2 I)\Lambda^2=O$$
1) \textbf{Assume $a^2>b^2$}.
$\Pi_\alpha$ is defined by $\mathbf{W}_1$ and $\mathbf{W}_2$ (\ref{W}) with $\omega=0$ :
\begin{eqnarray*}
\mathbf{W}_1& = & a^2 e_0+qC\\
\mathbf{W}_2 & = & \alpha^2\, qA
\end{eqnarray*}
and $$\langle \mathbf{W}_1,\mathbf{W}_1\rangle = -a^2\alpha^2,\quad\langle \mathbf{W}_2,\mathbf{W}_2\rangle = a^2\alpha^4$$
$\Pi_0$ is the kernel of $\Lambda$. It is generated by $\mathbf{V}_1$ (first column of $\Lambda^2-\alpha^2I$) and $\mathbf{V}_2=qB$ 
\begin{eqnarray*}
\mathbf{V}_1 & = & (a^2-\alpha^2)\,e_0+qC =b^2\,e_0+qC\\
\mathbf{V}_2 & = & qB 
\end{eqnarray*}
As above, normalizing the four vectors and writing them $(E_0,E_1,E_2,E_3)$ respectively, the reduced form of the matrix $\Lambda$ in this new basis shall be the first matrix of (\ref{fr2})\\
2) \textbf{Assume} $a^2<b^2$ :
Noting $\omega^2=b^2-a^2$ we have
$$Pm(\Lambda^2)=(\Lambda^2+\omega^2 I)\Lambda^2=O$$
$\Pi_0$ is timelike since it is generated by the two vectors belonging to the kernel of $\Lambda$:
\begin{eqnarray*}
\mathbf{W}_1&=&(a^2+\omega^2)e_0+qC=b^2e_0+qC \\
\mathbf{W}_2&=&qB 
\end{eqnarray*}
$$\langle \mathbf{W}_1,\mathbf{W}_1\rangle = -\omega^2a^2,\quad\langle \mathbf{W}_2,\mathbf{W}_2\rangle = \omega^4a^2$$
The plane $\Pi_\omega$ is space like. It is generated the first column $\textbf{V}_1$ of $\Lambda^2$ and $\textbf{V}_2=\Lambda .\textbf{V1}$ :
\begin{eqnarray*}
\mathbf{V}_1&=&a^2e_0+qC \\
\mathbf{V}_2&=&\Lambda .\mathbf{V}_1=q(a^2A+B\times C)=q((a^2-b^2)A)\\
&=&-\omega^2qA
\end{eqnarray*}
$$\langle \mathbf{V}_1,\mathbf{V}_1\rangle = \omega^2a^2,\quad\langle \mathbf{V}_2,\mathbf{V}_2\rangle = \omega^4a^2$$
As above, normalizing  the four vectors (the first one being timelike) and writing them $(E_0,E_1,E_2,E_3)$, the reduced form of $\Lambda$ in this new basis shall be the second matrix of (\ref{fr2})\\
3) \textbf{Assume $a^2=b^2$} : when $a^2=b^2$ the minimal polynomial of $\Lambda^2$ is simply $Pm(X)=X^2$ that is to say: 
$$Pm(\Lambda^2)=\Lambda^2 .\Lambda^2=O$$\
Noting $$E_0=e_0,\quad E1=\frac{1}{a}qA,\quad E2=\frac{1}{a}qB,\quad E3=\frac{1}{a^2}qC$$ $(E_0,E_1,E_2,E_3)$ form an $\eta$-orthonormal basis and $(E_0,E_1)$ generate an eigenspace $\Pi_0$ with the relations:
\begin{eqnarray*}
\Lambda .E_0 & = & qA=aE_1 \\
\Lambda .E_1 & = & \frac{1}{a}\Lambda .qA=\frac{1}{a}(a^2e_0+q(B\times A))=aE_0+aE_3\\
\Lambda .E_2 & = & 0\\
\Lambda .E_3 & = & \frac{1}{a^2}\Lambda qC=\frac{1}{a^2}q(B\times C)=\frac{1}{a^2}(-b^2qA)=-aE_1
\end{eqnarray*}
We thus obtain the third reduced form in (\ref{fr2}).
\end{description}
\textbf{Corollary} :\\
\textit{The Lie group of special and orthochronous Lorentz matrices has four one-parameter subgroups which can be obtained by integrating the linear differential equation $$\frac{dS}{dt}=S.\Lambda$$ where $\Lambda$ is one of the four reduced forms obtained above.}\\
The solution of this equation is $S_0.e^{t\Lambda}$ that is to say \\
\begin{eqnarray*}
S_1(t) & = & S_0\left(\small
\begin{array}{cccc}
 \cosh \alpha t & \sinh \alpha t & 0 & 0 \\
 \sinh \alpha t & \cosh \alpha t & 0 & 0 \\
 0 & 0 & \cos \omega t & -\sin \omega t \\
 0 & 0 & \sin \omega t & \cos \omega t
\end{array}
\right)
\\
S_2(t) & = & S_0\left(\small
\begin{array}{cccc}
 \cosh \alpha t & \sinh \alpha t & 0 & 0 \\
 \sinh \alpha t & \cosh \alpha t & 0 & 0 \\
 0 & 0 & 1 & 0\\
 0 & 0 & 0 & 1
\end{array}
\right)
\\
S_3(t) & = & S_0\left(\small
\begin{array}{cccc}
 1 & 0 & 0 & 0 \\
 0 & 1 & 0 & 0 \\
 0 & 0 & \cos \omega t & -\sin \omega t \\
 0 & 0 & \sin \omega t & \cos \omega t
\end{array}
\right)
\\
S_4(t) & = &S_0 \left(\small
\begin{array}{cccc}
 \frac{1}{2} a^2 t^2+1 & a t & 0 & -\frac{1}{2} a^2 t^2 \\
 a t & 1 & 0 & -a t \\
 0 & 0 & 1 & 0 \\
 \frac{1}{2}a^2 t^2 & a t & 0 & 1-\frac{1}{2} a^2 t^2 \\
\end{array}
\right)
\end{eqnarray*}                                                                                                                                                                    
\section{Inertial particles in Special Relativity}
Let $O$ and $M$ be two inertial particles in the Minkowski space $(\mathcal{M},\eta)$.
Their worldlines $\mathcal{L}_O$ and $\mathcal{L}_M$ are two geodesic straight lines of $\mathcal{M}$ generated by the timelike future oriented unitary 4-vectors $\textbf{t}$ and by $\textbf{V}$ (which define the 4-velocities of $O$ and $M$ respectively). \\
Let $\mathcal{R}_O=(e_0=\textbf{t},e_1,e_2,e_3)$ be the $\eta$-orthonormal basis associated to $O$ along $\mathcal{L}_O$ and let us recall that $(e_1,e_2,e_3)$ is a basis of the hyperplane of $\mathcal{M}$ passing through $O$ and orthogonal to the worldline of $O$. This hyperplan is the physical space of $O$. 
We note $t$ and $\tau$ the proper times of $O$ and $M$ respectively. We also denote $(t,x,y,z)$ the coordinates of $M$ in the referential frame of $O$ and $V=(p,q,r)$ the 3-velocity of $M$. Using these notations, the 4-velocity $\textbf{V}=\displaystyle \frac{dM}{d\tau}$ can be written
\begin{eqnarray*}
\textbf{V} & = & \frac{dt}{d\tau}\,(1,p,q,r)\\
 & = & \Gamma \,e_0+\Gamma\, V 
\end{eqnarray*}
with the relations:
$$\langle \textbf{V}, \textbf{V}\rangle=-1\quad \Rightarrow \quad \Gamma^2(1-V^2)=1,\quad V^2=p^2+q^2+r^2$$
where $\Gamma=\displaystyle\frac{dt}{d\tau}$ is the Lorentz factor. All these quantities are constants.

In order to define the Lorentz-Poincaré transform we may apply the orthonormalization Gram-Schmidt process to the basis 
$(\textbf{V},e_1,e_2,e_3)$. We thus obtain an $\eta$-orthonormal basis, $(M, E_0,E_1,E_2,E_3)$ where $E_0=\textbf{V}$ and where the three other vectors generate the basis of the physical space of $M$. This orthonormalization process directly gives the boost characterizing the relation between the two inertial particles: 
\begin{equation}\label{boost}
L=\left(\small
\begin{array}{cc}
 \Gamma & \Gamma\,{^T}\!V   \\
\Gamma\, V &\,\,\, I_3+ \dfrac{\Gamma^2}{1+\Gamma}V\,{^T}\!V\\
 \end{array}
\right)
\end{equation}
In this result,  $V$ is the column matrix of its components and $I_3$ is the unit matrix of size $3$.

$L$ being a constant matrix, its associated matrix in the Lie algebra is the zero matrix. All this corresponds to the classical case of Special Relativity and can be summarized as follows: 
\textit{Any constant matrix $L \in \mathcal{S}$ defines a Lorentz transform relating two inertial particles}. \\
\textbf{Remarks:}
\begin{description}
  \item[1] The relation between $O$ and $M$ can be characterized by an infinity of Lorentz matrices. Each of them can be deduced from $L$ by a left or a right multiplication of $L$ with a  pure rotation (a Lorentz matrix) $R$
$$
R=\left(\small
\begin{array}{cc}
1 &0  \\
0 &A\\
 \end{array}
\right)
$$
where $A$ is an orthogonal matrix of size 3. A left and a right multiplication correspond to  a change of basis in the rest space of $O$ and of $M$ respectively.
  \item[2 ] The writing of the boost (\ref{boost}) can be simplified by choosing an appropriate basis of $(O,e)$ (recall that $e_0$ is the 4-velocity of $O$ and let us note $qV$ the 4-vector associated to the 3-velocity $V=(p, q, r)$ of $M$ in $(O,e)$). $e_0$ and $qV$ are two orthogonal vectors in the Lorentz-Poincaré transform plane. In fact, noting $v^2=p^2+q^2+r^2$ 
\begin{eqnarray*}
L.e_0 & = & \mathbf{V} =\Gamma e_0+\Gamma\, qV\\
L.qV & = & \Gamma v^2 e_0+\Gamma\, qV 
\end{eqnarray*} 
We can define an $\eta$-orthonormal basis $(e'_0,e'_1)$ of this timelike plane by taking $e'_0=e_0$ and $e'_1=\displaystyle\frac{1}{v}\,qV$. We thus obtain:
\begin{eqnarray*}
L.e'_0 & = & \Gamma\,e'_0+\Gamma v\,e'_1 \\
L.e'_1 & = &\Gamma v\,e'_0+\Gamma\,e'_1 
\end{eqnarray*}
We also know that the two dimensional orthogonal complement is L-invariant. This can be seen by noting that the two 4-vectors $W_1=(0,-q,p,0)$ and $W_2=(0,-r,0,p)$ are orthogonal to $e'_0$ and to $e'_1$ and that they are linearly independant so that they form a basis. We can then construct an orthonormal basis of the spacelike plane which remains unchanged when orthonormalization process is applied:
\begin{eqnarray*}
e'_2 & = & \frac{1}{\sqrt{p^2+q^2}} (0,\,-q,\,p,\,0)\\
e'_3 & = & \left(0,\,-\frac{p\, r}{v \sqrt{p^2+q^2}},\,-\frac{q\, r}{v
   \sqrt{p^2+q^2}},\,\frac{\sqrt{p^2+q^2}}{v}\right)
\end{eqnarray*} 
These two vectors are eigenvectors of $L$ associated to the double eigenvalue $1$. We thus obtain a new $\eta$-orthonormal basis $(O,e'_0,e'_1,e'_2,e'_3)$ the transfert matrix beeing the Lorentz matrix $Q$:
\begin{equation*}
\label{Q}
Q=\left(\small
\begin{array}{cccc}
 1 & 0 & 0 & 0 \\
 0 & \frac{p}{v} & -\frac{q}{\sqrt{p^2+q^2}} & -\frac{p r}{\sqrt{p^2+q^2}
   v} \\
 0 & \frac{q}{v} & \frac{p}{\sqrt{p^2+q^2}} & -\frac{q r}{\sqrt{p^2+q^2}
   v} \\
 0 & \frac{r}{v} & 0 & \frac{\sqrt{p^2+q^2}}{v} \\
\end{array}
\right)
\end{equation*}
$Q$ is a pure rotation matrix ($^T\! Q.Q=I$) which only depends on the velocity direction.\\.\\
Noting
$$V_e=(v \cos (\alpha ) \cos (\beta ),\,v \cos (\alpha ) \sin (\beta ),\,v \sin (\alpha ))$$ the above expression can also be written
\begin{equation}
\label{Q}
Q=
\left(\small
\begin{array}{cccc}
 1 & 0 & 0 & 0 \\
 0 & \cos (\alpha ) \cos (\beta ) & -\sin (\beta ) & -\cos (\beta ) \sin (\alpha
   ) \\
 0 & \cos (\alpha ) \sin (\beta ) & \cos (\beta ) & -\sin (\alpha ) \sin (\beta
   ) \\
 0 & \sin (\alpha ) & 0 & \cos (\alpha ) \\
\end{array}
\right)
\end{equation}
To summarize : there is a basis $e'$ deduced from $e$ through a space rotation of $e$ for which the boost $L$ can be written in the following canonical form:
\begin{equation}
\label{Lcan}
L_{e'}=Q^{-1}.L.Q=\left(\small
\begin{array}{cccc}
 \Gamma  & v \Gamma  & 0 & 0 \\
 v \Gamma  & \Gamma  & 0 & 0 \\
 0 & 0 & 1 & 0 \\
 0 & 0 & 0 & 1 \\
\end{array}
\right)
\end{equation}
With respect to $e'$, $(M,E_0,E_1)$ is the plane of the Lorentz transformation and $(M,E_2,E_3)$ is the invariant plane of that transformation.

\end{description}
\section{\textit{Non inertial} particles in Special Relativity. Tangent Boost along a worldline.}
Let us now consider the case where $O$ is an inertial particle and where $M$ is not. Then, the wordline $\mathcal{L}_M$ of $M$ is no more a straight line and its 4-velocity $\textbf{V}$ is a vector field along $\mathcal{L}_M$. This leeds us to define \textit{the tangent boost along $\mathcal{L}_M$ as being the boost of the inertial particle $M'$ which coincides with $M$ and the worldine of which is the tangent straight line at $M$}. We thus obtain a field $\tau \rightarrow L(\tau)$ along $\mathcal{L}_M$ where $L$ is defined by (\ref{boost}). $L$ being no more a constant matrix, its associated matrix in the Lie algebra $\Lambda =L^{-1}\displaystyle\frac{dL}{d\tau}$ is no more the zero matrix. Before computing the 3-vectors $A$ and $B$ of the $\Lambda (A,B)$ matrix, let us give some examples of using this latter.

\subsection{Derivation rule of a vector $X$ defined by its components in the referential frame of $M$}
Let us consider the two basis $e=(e_0,e_1e_2,e_3)$ and $E=(E_0,E_1,E_2,E_3)$, $E$ being defined by the columns of $L$.
Let $X_e$ and $X_E$ be the components of the $\mathbf{X}$ vector in $e$ and $E$ respectively ($X_e=L.X_E$). Let us derivate that relation with respect to $t$ (or with respect to the proper time $\tau$ of $M$). Using then the left translation $\mathcal{L}_{L^{-1}}$, we get:
\begin{eqnarray*}
L^{-1} \frac{dX_e}{dt} & = & \frac{dX_E}{dt} +L^{-1}.\frac{dL}{dt} X_E 
 =  \frac{dX_E}{dt} +\Lambda_E .\,X_E 
\end{eqnarray*}
where the subscripts $e$ and $E$ correspond to the basis $e$ and $E$ respectively.
The above relation gives \textit{the derivative rule} by its $E-components$ that is the intrinsic vectorial relation:
\begin{equation}\label{derivation}
\left(\frac{d\mathbf{X}}{dt}\right)_e= \left(\frac{d\mathbf{X}}{dt}\right)_E + \mathbf{\Lambda}.\mathbf{X}
\end{equation}
Let us now apply that law to the $4$-velocity of $M$ the components of which are $(1,0,0,0)$ in $E$. \\ Eq.(\ref{derivation}) shows that the first column of $\Lambda(A,B)$ is the 4-acceleration $qA=(0,a_1,a_2,a_3)$ (notation n2 in paragraph 3.3) of $M$ in $E$.\\
Now, let $\textbf{W}$ be a 4-vector defined by its components $W_E=(w_0,w_1,w_2,w_3)=w_0 \,E_0+qW$ in $E$ and let us recall that  $E_0=\mathbf{V}$ is the 4-velocity of $M$. Noting $W$ the 3-vector $(w_1,w_2,w_3)$,  and $\mathbf{A}$ the 4-acceleration of $M$ with $A_E=qA$, 
the 3-vector $B=(b_1,b_2,b_3)$ appears to be an instantaneous rotation defined by its components in $E$:
\begin{eqnarray*}
\frac{dW_e}{d\tau} & = & \frac{dW_E}{d\tau}+\Lambda_E.W_E\nonumber \\
& = &  \frac{dW_E}{d\tau}+(W.A) E_0+q(w_0 A+B\times W)\nonumber \\
& = & \frac{dW_E}{d\tau}+\langle W_E,qA\rangle E_0+q(w_0 A+B\times W)\nonumber \\
 \left(\frac{d\mathbf{W}}{d\tau}\right)_e & = & \left(\frac{d\mathbf{W}}{d\tau}\right)_E+\langle \mathbf{W},\mathbf{A}\rangle \mathbf{V}+w_0 \mathbf{A}+q(B\times W)
\end{eqnarray*}
Changing $w_0$ into $- \langle \mathbf{V},\,\mathbf{W} \rangle$ this last equation can also be written:
\begin{equation}\label{regl} 
 \left(\frac{d\mathbf{W}}{d\tau}\right)_e  =  \left(\frac{d\mathbf{W}}{d\tau}\right)_E+\langle \mathbf{W},\mathbf{A}\rangle \mathbf{V} - \langle \mathbf{V},\,\mathbf{W} \rangle \mathbf{A}+q(B\times W)
\end{equation}
Note that there is a minor abuse of notation in the last line: $B$ and $W$ must be understood here as 3-vectors and no more as components in $E$ as in previous lines. 
The term $q(B\times W)$ shows that $B$ is an instantaneous rotation in the (physical) space of 3-vectors. It corresponds to Thomas rotation.\\
The matrix $\Lambda$ thus contains the 4-acceleration of $M$ and the Thomas rotation. It therefore undoubtedly constitutes a valuable tool to describe the motion of any physical system.

\subsection{Example of an uniformly accelerated particle}
In the referential frame $\mathcal{R}_O$ of an inertial observer $O$, an uniform acceleration of $M$ does not correspond to a constant 4-acceleration $\mathbf{A}$. In fact, the worldline $\mathcal{L}_M$ of $M$ is not a straigth line since it is not a geodesic. At two different points $M_1$ and $M_2$, $\mathbf{A}_1$ and $\mathbf{A}_2$ are not parallel. In the case of an uniformly accelerated particule, we consequently only know that the norm of  $\mathbf{A}$ is a constant $a$. Moreover, in what follows, we will also consider that, for the inertial observer $O$, $\mathcal{L}_M$ remains in a given plane  $(e_0,e_1)$. This plane is necessarily a timelike plane. The parametric equation of motion for $M$ and its 4-velocity $\mathbf{V}=\displaystyle\frac{dM}{d\tau}$ are then:
\begin{eqnarray*}
M(\tau) & = & \left( t(\tau),\,x(\tau),\,0,\,0\right) \\
\mathbf{V} & = & \Gamma (1,\,v,\,0,\,0),\qquad v=\frac{dx}{dt},\qquad \Gamma^2(1-v^2)=1
\end{eqnarray*}
where:
\begin{equation}
\label{rel5}
\Gamma^2(1-v^2)=1\quad\Rightarrow\quad \frac{d\Gamma}{d\tau}=\Gamma^3\,v\,\frac{dv}{d\tau},\qquad \frac{d(v\Gamma)}{d\tau}=\Gamma^3\,\frac{dv}{d\tau}
\end{equation}
The mere knowledge of $\mathbf{V}$ permits to calcule the tangent boost $L$. Inserting $V=(v,0,0)$ into eq.(\ref{boost}) we get  :
\begin{equation}\label{}
L=\left(\small
\begin{array}{cccc}
 \Gamma & \Gamma\,v & 0 & 0   \\
\Gamma\, v & \Gamma & 0 & 0  \\
0 & 0 & 1 & 0\\
0 & 0 & 0 & 1
 \end{array}
\right)
\end{equation}
Let us then calculate its associated matrix $\Lambda$ in the Lie algebra
$$\Lambda=L^{-1}.\frac{dL}{d\tau}=\eta .^T\! L.\eta .\frac{dL}{d\tau}$$
Using (\ref{rel5}) in computing $\displaystyle\frac{dL}{d\tau}$, the above equation gives:
\begin{equation}\label{}
\Lambda=\Gamma^2\frac{dv}{d\tau}\left(\small
\begin{array}{cccc}
0 & 1 & 0 & 0   \\
1 & 0 & 0 & 0  \\
0 & 0 & 0 & 0\\
0 & 0 & 0 & 0
 \end{array}
\right)
=\left(\small
\begin{array}{cccc}
0 & a & 0 & 0   \\
a & 0 & 0 & 0  \\
0 & 0 & 0 & 0\\
0 & 0 & 0 & 0
 \end{array}
\right)
\end{equation}
where $\displaystyle\Gamma^2\frac{dv}{d\tau}=a$ is the constant defined above (when $\displaystyle\frac{dv}{d\tau}>0$, $a$ is the norm of the 4-acceleration).
Using the derivation rule, we obtain the components $\mathbf{A}_E$ of the 4-acceleration in $(M,E)$
$$\mathbf{A}_E=\Lambda.\mathbf{V}_E=(0,\,\Gamma^2\frac{dv}{d\tau},\,0,\,0)$$
Its components $\mathbf{A}_e$ in $(O,e)$ are then obtained by a change of basis
$$\mathbf{A}_e=L.\mathbf{A}_E=\Gamma^3\frac{dv}{d\tau}(v,\,1,\,0,\,0)$$
Calculating $ L(\tau)=e^{\tau\Lambda}$ we get the following conclusions: \textit{any uniformly accelerated particle is defined by a one-parameter subgroup of the Lie group $\mathcal{S}$}, 
$$\tau\rightarrow L(\tau)=e^{\tau\Lambda}=\left(\small
\begin{array}{cccc}
 \cosh (a \tau ) & \sinh (a \tau ) & 0 & 0 \\
 \sinh (a \tau ) & \cosh (a \tau ) & 0 & 0 \\
 0 & 0 & 1 & 0 \\
 0 & 0 & 0 & 1 \\
\end{array}
\right)$$
\textit{and the 4-acceleration is uniform in the rest frame $(M,E)$ of $M$ (note again that the basis $E$ is defined by the columns of $L$)}. In an uniformly accelerated system, there is no Thomas rotation.\\
Let us now consider two nearby particles $N$ and $M$, $N$ being at rest with respect to $M$ and their coordinates in
 $(M,E)$ being $MN=(0,X,0,0)$ where $X= constant$. Let us calculate  $\displaystyle\frac{dON}{d\tau}$. $$\frac{dON}{d\tau}=\frac{dOM}{d\tau}+\frac{dMN}{d\tau}=\textbf{V}_M+\frac{dMN}{d\tau}$$
Knowing that $X$ does not depend on $\tau$, the derivation rule gives:
$$\frac{dMN}{d\tau}=\Lambda.MN=(aX,\,0,\,0,\,0)$$
The components of $\displaystyle\frac{dON}{d\tau}$ in $(M,E)$ are then
$$\frac{dON}{d\tau}=(1+aX,\,0,\,0,\,0)$$
where $1+aX$ is a velocity (using $ c\neq 1$ we would get $\displaystyle\ c+\frac{aX}{c}$ instead of $1+aX$). It is important to note that $\displaystyle\frac{dON}{d\tau}$ is not the 4-velocity of $N$ and that the proper time of $N$ is not the same as the one of $M$. In fact, the norm $\textbf{V}_N$ of the 4-velocity of $N$, defined with its proper time $s$ being 1, we obtain the following relation between $\tau$ and $s$ :
$$\textbf{V}_N=\frac{dON}{ds}=\frac{dON}{d\tau}\frac{d\tau}{ds}=(1,\,0,\,0,\,0)\quad \Rightarrow\quad \frac{d\tau}{ds}=\frac{1}{1+aX}$$
This shows that \textit{in the case of a non inertial motion of $M$, it is impossible to synchronize the clocks in the rest frame of $M$}.\\
Let us add that $N$ has not the same acceleration as $M$. In fact, knowing that $\textbf{V}_N=\textbf{V}_M$ and consequently that $\displaystyle\frac{d\textbf{V}_N}{d\tau}=\frac{d\textbf{V}_M}{d\tau}=\textbf{A}_M$, we get
$$\textbf{A}_N=\frac{d\textbf{V}_N}{ds}=\frac{d\textbf{V}_N}{d\tau}\frac{d\tau}{ds}=\frac{1}{1+aX}\frac{d\textbf{V}_N}{d\tau}=\frac{1}{1+aX}\textbf{A}_M$$

\subsection{Tangent boost of a worldline and its associated matrix in the Lie algebra  in Special Relativity}
In the referential frame of $O$, the parametric equations of the worldline $\mathcal{L}_M$ are defined by cartesian coordinates where the parameter is the proper time $\tau$ of $M$ :
$$\tau\rightarrow M(\tau)=(t(\tau),\,x(\tau),\,y(\tau),\,z(\tau))$$
Noting $V_e=(x',y',z')$ and $A_e=(x'',y'',z'')$ the 3-velocity and the 3-acceleration in the reference frame of $O$ (with its propertime $t$) the 4-velocity and the 4-acceleration (first and second derivative of coordinates with respect to $t$) are:
\begin{eqnarray*}
\mathbf{V} & = & \Gamma(1,x',y',z')=\Gamma (e_0+V_e);\qquad \Gamma^2(1-V_e^2)=1 \\
\mathbf{A} & = & \frac{d\Gamma}{d\tau}(e_0+V_e)+\Gamma^2 qA_e
\end{eqnarray*}
$$\Gamma^2(1-V_e^2)=1\quad \Rightarrow \quad\frac{d\Gamma}{d\tau}=\Gamma^4(V_e.A_e)
$$
The tangent boost (\ref{boost}) is:
\begin{equation}\label{bt}
L(\tau)=\left(\small
\begin{array}{cccc}
 \Gamma  & \Gamma  x' & \Gamma  y' & \Gamma  z' \\
 \Gamma  x' & \frac{\Gamma ^2 \left(x'\right)^2}{\Gamma +1}+1 & \frac{\Gamma ^2 x'
   y'}{\Gamma +1} & \frac{\Gamma ^2 x' z'}{\Gamma +1} \\
 \Gamma  y' & \frac{\Gamma ^2 x' y'}{\Gamma +1} & \frac{\Gamma ^2 \left(y'\right)^2}{\Gamma
   +1}+1 & \frac{\Gamma ^2 y' z'}{\Gamma +1} \\
 \Gamma  z' & \frac{\Gamma ^2 x' z'}{\Gamma +1} & \frac{\Gamma ^2 y' z'}{\Gamma +1} &
   \frac{\Gamma ^2 \left(z'\right)^2}{\Gamma +1}+1 \\
\end{array}
\right)
\end{equation}
and its associated matrix in the Lie algebra $\Lambda=L^{-1}\displaystyle\frac{dL}{d\tau}$ is
\begin{equation*}\label{lb}
\Lambda=\left(\small
\begin{array}{cccc}
 0 & \frac{(V_e.A_e) x' \Gamma ^4}{\Gamma +1}+x'' \Gamma ^2 & \frac{(V_e.A_e) y' \Gamma ^4}{\Gamma
   +1}+y'' \Gamma ^2 & \frac{(V_e.A_e) z' \Gamma ^4}{\Gamma +1}+z'' \Gamma ^2 \\
 \frac{(V_e.A_e)
 x' \Gamma ^4}{\Gamma +1}+x'' \Gamma ^2 & 0 & \frac{\Gamma ^3 \left(y' x''-x'
   y''\right)}{\Gamma +1} & \frac{\Gamma ^3 \left(z' x''-x' z''\right)}{\Gamma +1} \\
 \frac{(V_e.A_e) y' \Gamma ^4}{\Gamma +1}+y'' \Gamma ^2 & \frac{\Gamma ^3 \left(x' y''-y'
   x''\right)}{\Gamma +1} & 0 & \frac{\Gamma ^3 \left(z' y''-y' z''\right)}{\Gamma +1} \\
 \frac{(V_e.A_e) z' \Gamma ^4}{\Gamma +1}+z'' \Gamma ^2 & \frac{\Gamma ^3 \left(x' z''-z'
   x''\right)}{\Gamma +1} & \frac{\Gamma ^3 \left(y' z''-z' y''\right)}{\Gamma +1} & 0 \\
\end{array}
\right)
\end{equation*}
To summarize: using notations (\ref{lambda}) we see that $\Lambda$ gives the complete dynamics of $M$. In $\Lambda(A,B)$ :
\begin{itemize}
  \item  the 3-vector $A$ is the acceleration of $M$ in its rest frame $(M, E_0,E_1,E_2,E_3)$:
 \begin{equation}
\label{Ac}
A=\displaystyle\frac{\Gamma^4}{\Gamma +1}(V_e.A_e)\, V_e+\Gamma^2A_e
\end{equation}
 \item  the 3-vector $B$ gives the instantaneous Thomas rotation by its components in $(M, E_0,E_1,E_2,E_3)$: 
\begin{equation}
\label{Thomas}
B=\Omega_T=\displaystyle\frac{\Gamma^3}{\Gamma +1}V_e\times A_e
\end{equation}
\end{itemize}
\subsubsection{Writing the tangent boost and its associated matrix in the Lie algebra in a rotating frame $(O,e')$. A first insight on Thomas rotation}
The rotating basis $e'$ is defined in the remark (2) of  paragraph 4 but, in the present case, the rotation matrix $Q$ now depends on the proper time of M. The tangent boost $L$ in $e'$ has the remarkable form (\ref{Lcan}). 
Our aim is to calculate the components of the matrix of the Lie algebra in the rotating frame $(O,e')$ in two ways:
\begin{description}
  \item[1 ]  Using the definition of $\Lambda$ in the moving referential frame $e'$
\begin{eqnarray*}
\Lambda_{e'}=Q^{-1}.\Lambda.Q  & = & \,^T\!Q.L^{-1}.\frac{dL}{d\tau}.Q =  L_{e'}^{-1} \,^T\!Q.\frac{dL}{d\tau}.Q
\end{eqnarray*}  
and applying the derivation rule to the tangent boost $L_{e'}$ in $e'$ 
\begin{equation*}
^T\!Q.\frac{dL}{d\tau}.Q  =  \,^T\!Q.\frac{d}{d\tau}(Q.L_{e'}\,^T\!Q).Q 
 = \frac{dL_{e'}}{d\tau} +\Omega.L_{e'}-L_{e'}.\Omega
\end{equation*} 
where $\displaystyle\Omega=\,^T\!Q.\frac{dQ}{d\tau}$ is the antisymmetric matrix which defines the instantaneous rotation of $(O,e')$. Inserting this result in the previous equation gives the matrix of the Lie algebra of the boost $L_{e'}$ as seen by the rotating observer $(O,e')$
 $$\Lambda_{e'}=L_{e'}^{-1}.\frac{dL_{e'}}{d\tau}+L_{e'}^{-1}.\Omega.L_{e'}-\Omega$$ 
We thus obtain $\Omega$ and $\Lambda_{e'}$
\begin{equation*}
\label{ }
\Omega=\,^T\!Q.\frac{dQ}{d\tau}=
\left(\small
\begin{array}{cccc}
 0 & 0 & 0 & 0 \\
 0 & 0 & -\Gamma  \cos (\alpha ) \beta ' &
   -\Gamma  \alpha ' \\
 0 & \Gamma  \cos (\alpha ) \beta ' & 0 & -\Gamma
    \sin (\alpha ) \beta ' \\
 0 & \Gamma  \alpha ' & \Gamma  \sin (\alpha )
   \beta ' & 0 \\
\end{array}
\right)
\end{equation*}
\begin{equation*}
\label{ }
\Lambda_{e'}=
\left(\small 
\begin{array}{cccc}
 0 & \Gamma ^3 v' & v \Gamma ^2 \cos (\alpha ) \beta ' & v \Gamma ^2 \alpha ' \\
 \Gamma ^3 v' & 0 & -\frac{v^2 \Gamma ^3 \cos (\alpha ) \beta '}{\Gamma +1} &
   -\frac{v^2 \Gamma ^3 \alpha '}{\Gamma +1} \\
 v \Gamma ^2 \cos (\alpha ) \beta ' & \frac{v^2 \Gamma ^3 \cos (\alpha ) \beta
   '}{\Gamma +1} & 0 & 0 \\
 v \Gamma ^2 \alpha ' & \frac{v^2 \Gamma ^3 \alpha '}{\Gamma +1} & 0 & 0 \\
\end{array}
\right)
\end{equation*}
where $\alpha', \beta'$ and $v'$ are the derivatives with respect to $t$ of the three parameter defining $V$ (let us recall that $\displaystyle\frac{d}{d\tau}=\Gamma \frac{d}{dt}$).
  \item[2 ]  Using (\ref{Ac}) and (\ref{Thomas}) which give the 3-vectors A and B from $V_e$ and $A_e$.
we get:
\begin{equation*}
\label{ }
A_e=
\left(\small
\begin{array}{c}
\cos (\alpha ) \cos (\beta ) v'-v \alpha ' \sin (\alpha ) \cos (\beta )-v\cos (\alpha ) \beta ' \sin (\beta ) \\
\cos (\alpha ) \sin (\beta ) v'-v \alpha ' \sin (\alpha ) \sin (\beta )+v \cos (\alpha ) \beta ' \cos (\beta ) \\
 \sin(\alpha ) v'+v \alpha ' \cos (\alpha ) 
\end{array}
\right)   
\end{equation*}
 \begin{eqnarray*}
V_{e'} & = & \,^T\!Q.V_e=(v,0,0) \\
A_{e'} & = &  \,^T\!Q.A_e =(v',\,v \cos (\alpha ) \beta ',\,v \alpha ')
\end{eqnarray*}
 Using $\Gamma^2v^2=\Gamma^2-1$,  eq.(\ref{Ac}) gives :
\begin{eqnarray*}
A & = & \frac{\Gamma^4}{\Gamma +1}(V_{e'}.A_{e'})\, V_{e'}+\Gamma^2A_{e'} \\
 & = &\Gamma^2\left(\Gamma  v', \,v \cos (\alpha ) \beta ',\, v \alpha'\right)
   \end{eqnarray*}
and eq. (\ref{Thomas}) gives the Thomas rotation $B=\Omega_T$ :
\begin{eqnarray*}
\Omega_T & = & \frac{\Gamma^3}{\Gamma +1}V_{e'}\times A_{e'}\\
 & = & \frac{\Gamma^3}{\Gamma +1}\left(0,\,-v^2\alpha',\,v^2\cos(\alpha)\beta'\right) 
\end{eqnarray*} 
  \end{description}
To conclude: from the $(O,e)$ observer point of view, the L boost written in the rotating basis $e'$ defines the rest frame $(M,e)$ with respect to $(O,e')$. From the $(O,e')$ point of view, the two dimensional space of the Lorentz-Poincaré transform, as well as its invariant space are not moving. The matrix $Q$ defines the rotation of the rest frame of $M$ with respect to $(O,e)$.\\
\textit{These calculations show that we must clearly distinguish between the instantaneous rotation of $(O,e')$ (which is defined from the antisymmetric matrix \\ $\displaystyle\Omega_{e'}= \,^T\!Q.\frac{dQ}{d\tau}$) and the instantaneous Thomas rotation}.\\

In order to get a better insight on Thomas rotation, let us consider the infinitesimal Lorentz matrix relating $M(\tau)$ to $M(\tau+d\tau)$ :
\begin{eqnarray*}
L(\tau+d\tau)&\simeq&L(\tau)\left(I+L^{-1}(\tau)\,\frac{dL}{d\tau} d\tau \right)\\
&\simeq&L(\tau)\left(I+\Lambda(\tau) d\tau\right)
\end{eqnarray*}
A left-multiplication by $L^{-1}$ of this result gives the Lorentz matrix 
\begin{eqnarray*}
Li & = & L^{-1}(\tau)L(\tau+d\tau) = I+\Lambda(\tau)\,d\tau\nonumber +o(d\tau) \\
   & = & I+(\Lambda _A +\Lambda _B) d\tau+o(d\tau)\\
   & = & (I+\Lambda_Ad\tau)(I+\Lambda_Bd\tau) +o(d\tau)\
\end{eqnarray*}
At first order, $Li$ thus appears to be the product of an infinitesimal boost $Bi=I+\Lambda_Ad\tau$ with an infinitesimal pure rotation (Thomas rotation) $Ri=I+\Lambda_Bd\tau$  :
$$Li=Bi.Ri+o(d\tau)$$
We will see later that the Thomas rotation is a rotation of the \textit{rest frame of $M$} with respect to the referential frame $(M,E)$ which is defined by the tangent boost.
\subsubsection{Application to a particle in circular motion at constant velocity}
With respect to the frame 
 $(O,\,\partial_t,\,\partial_r,\,\partial_\theta,\,\partial_z)$ of O, the parametric equations of the particule worldline are those of a circular helix with axis $(O,\partial_t)$. Using cylindrical coordinates $(t,r,\theta,z)$
$$t\rightarrow M(t)=(t,\,R ,\,\omega t,\,0)$$ 
where $R$ and $\omega$ are constant and where $t$ is a function of $\tau$.The 4-velocity is:
\begin{eqnarray*}
\mathbf{V} & = & \Gamma\frac{dM}{dt}=\Gamma \left(1,\, 0,\, \omega ,\,0\right) \\
\Gamma & = & \frac{dt}{d\tau} =\frac{1}{\sqrt{1-R^2\omega^2}}
\end{eqnarray*}
Noting that the Lorentz factor $\Gamma$ is constant, the 4-acceleration is:
$$\mathbf{A}=\frac{d\mathbf{V}}{d\tau}=D_\mathbf{V} \mathbf{V}=(0,\,-R\,\Gamma^2\omega^2,\,0,\,0)$$
where $D_\mathbf{V} $ is the covariant derivative in the direction of
$\mathbf{V}$ expressed in cylindrical coordinate.
In order to calculate the tangent boost we have to express $\mathbf{V}$ in the $\eta-$orthonormal system $(M,\ e_0,\, e_1,\, e_2,\, e_3)$ obtained by applying the Gram-Schmidt orthonormalisation process to the natural basis $(\partial)=(\partial_t,\,\partial_r,\,\partial_\theta,\,\partial_z)$ and starting with the 4-vector $\partial_t$. In the present case, calculations are very simple since the basis  $(\partial)$ is already $\eta-$orthogonal. We get $\mathbf{V}=(\Gamma,\, 0,\, \Gamma R\omega,\, 0)$.
The tangent boost is then defined by using (\ref{boost}):
\begin{equation*}
 \label{ }
\mathbf{B}=\left(\small
\begin{array}{cccc}
 \Gamma  & 0 & R \Gamma  \omega  & 0 \\
 0 & 1 & 0 & 0 \\
 R \Gamma  \omega  & 0 & \Gamma  & 0 \\
 0 & 0 & 0 & 1 \\
\end{array}
\right)
\end{equation*}
Let us recall that the $\mathbf{B}$-columns give the referential frame $(E)=(E_0=\mathbf{V},\, E_1,\, E_2,\ E_3)$ of   $M$, 
and that the 4-vectors $E_\alpha$ are defined from their components in $(e)$. Let us also note that $(M,\, E_0,\, E_2)$ is the plane of the Poincaré-Lorentz transform, $(M,\, E_1,\, E_3)$ being the invariant orthogonal supplementary plane of the transformation.

The matrix of the Lie algebra $\mathbf{\Lambda}=\mathbf{B}^{-1}\displaystyle\frac{d\mathbf{B}}{d\tau}=\mathbf{B}^{-1}D_\mathbf{V}\mathbf{B}$ is
\begin{equation}
\label{Thc}
\mathbf{\Lambda}=\left(\small
\begin{array}{cccc}
 0 & -R \Gamma ^2 \omega ^2 & 0 & 0 \\
 -R \Gamma ^2 \omega ^2 & 0 & -(\Gamma -1) \Gamma  \omega  & 0 \\
 0 & (\Gamma -1) \Gamma  \omega  & 0 & 0 \\
 0 & 0 & 0 & 0 \\
\end{array}
\right)
\end{equation}
It directly gives the 3-acceleration and the instantaneous Thomas rotation (which both are in the physical space of $M$). 
Let us note that it is also possible to obtain the 3-vectors $A$ and $B$ of $\Lambda(A,B)$ from eqs.(\ref{Ac}) and (\ref{Thomas}): using $V_e=(0,\, R\omega,\,0)$ and 
$A_e=(-R\omega^2,\, 0,\, 0)$ we in fact obtain the 3-acceleration and the instantaneous Thomas rotation in $E$ ($E$ is defined by the column vectors of $L$):
\begin{eqnarray*}
A & = & \Gamma^2 A_e = (-\Gamma^2 R\omega^2,\, 0,\, 0) \\
B & = & \Omega_{Th}= \frac{\Gamma^3}{\Gamma +1}V_e\times A_e=(0,\,0,\,\Gamma(\Gamma-1)\omega)
\end{eqnarray*}
\subsubsection*{discussion}
In order to understand the meaning of Thomas rotation, let us consider a gyroscope and let us recall the definition of a gyroscopic torque along a worldline as given by Straumann \cite{Misner}\cite{Straumann}\cite{Ryder}:\\
A gyroscopic torque along a worldline $\mathcal{L}$ the 4-velocity and the proper time of which are \textbf{V} and $\tau$ respectively is a 4-vector \textbf{G} defined along $\mathcal{L}$, orthogonal to \textbf{V} and such that its derivative with respect to $\tau$ is proportional to \textbf{V}, that is to say :
\begin{equation}
\label{gyro}
\langle\mathbf{G},\mathbf{V}\rangle=0\qquad and \qquad\frac{d\mathbf{G}}{d\tau}=k \mathbf{V}
\end{equation}
These relations permit to calculate $k$. Noting $\mathbf{A}$ the 4-acceleration $\displaystyle\frac{d\mathbf{V}}{d\tau}$, we in fact get : 
$$\frac{d}{d\tau}\langle\mathbf{G},\mathbf{V}\rangle=0\,\Rightarrow \, \langle k \mathbf{V},\mathbf{V}\rangle+\langle\mathbf{G},\mathbf{A\rangle =0}\,\Rightarrow \, k=\langle\mathbf{G},\mathbf{A}\rangle$$
The
proportionality condition implies that the 4-vector $\mathbf{G}$ (which belongs to the physical space of $M$ along the worldline  
$\mathcal{L}$) \textit{rotates in that space.} In fact, let us write the differential equation
 (\ref{gyro}) with respect to the components of $\mathbf{G}$ in $(M,E)$. Noting $\mathbf{G}_E=(0,\,G_1,\,G_2,\,G_3)$ in $(M,E)$, the components of $\mathbf{G}$ in the inertial frame are then defined by : $$\mathbf{G}_e=\mathbf{B}.\,\mathbf{G}_E=(\Gamma R\omega G_2,\,G_1,\,\Gamma G_2,G_3)$$ In the inertial referential frame, eq.(\ref{gyro}) thus becomes:
$$\frac{d({\mathbf{B}.\mathbf{G}_E})}{d\tau}=k \mathbf{V}_e  \quad \quad \quad	\mathbf{V}_e= (\Gamma,\, 0,\, R \Gamma \omega,\,0)$$
Using the covariant derivative in cylindrical coordinates and noting derivatives with respect to $\tau$ by accentuated characters we get:
$$\frac{d(\mathbf{\mathbf{B}.G_E})}{d\tau}=D_\mathbf{V}(\mathbf{B}.\mathbf{G}_E)=(\Gamma R \omega G'_2,\,-\Gamma^2\,\omega\,G_2+G'_1,\,\Gamma(\omega G_1+G'_2),\,G'_3)$$
Identifying this result with $k\mathbf{V}_e=(k\,\Gamma,\,0,\,k\,\Gamma R\omega,\,0)$ gives  $k=R\omega G'_2$ and the three differential equations (note that  $1-R^2\omega^2 =\frac{1}{\Gamma^2}$):
\begin{eqnarray*}
G'_1 & = & \Gamma^2 \omega G_2 \\
G'_2 & = &- \Gamma^2 \omega G_1\\
G'_3 & = &0
\end{eqnarray*}
Taking initial conditions $\mathbf{G}_E(\tau=0)=(0,\,A_1,\,A_2,\,A_3)$, the solutions of these differential equations are 
$$\mathbf{G}_E(\tau)=(0,\,A_1 \cos(\Gamma^2\omega\tau)+A_2\sin(\Gamma^2\omega\tau),\,A_2 \cos(\Gamma^2\omega\tau) - A_1\sin(\Gamma^2\omega\tau),\,A_3)$$
Fig.1 shows the rotation of a gyroscope initially oriented following the $x$ axis ($\mathbf{G}_E(0)=(0,\,1,\,0,\,0)$), in $(M,E)$ when $\tau$ varies from $0$ to $\frac{\pi}{\Gamma\omega}$ that is to say when $M$ goes a 180 degree turn. It shows that in that case, the gyroscope indicates a half turn plus a rotation which corresponds to the Thomas rotation (in the clockwise direction)\\
\begin{figure}[h]
\centering
\includegraphics[scale=0.6]{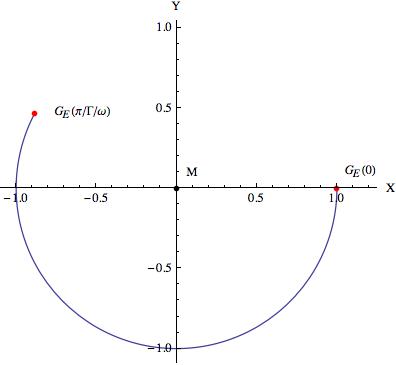}
\caption{\scriptsize{The gyroscope rotation in the plane $(M,E_1,E_2)$ when  $M$ goes a 180 degree turn.  Numerical values are $R=1$ and $\omega=\frac{5}{10}$}}
\end{figure}
It is also possible to highlight the Thomas rotation by applying the derivation rule (\ref{derivation}) to $\mathbf{G}_E$ (which is defined by its components in $(M,E)$). Noting $$\mathbf{G}_E=(0,\vec{G}); \quad \vec{G}=(G_1,\,G_2,\,G_3);\quad  \Omega_E=(0,\,\vec{\Omega}_{Th});\quad \vec{\Omega}_{Th}=(0,\,0,\,\Gamma(\Gamma-1)\omega)$$ and using (\ref{derivation}) in that moving frame:
$$D_\mathbf{V}\mathbf{G}= \left(\frac{d\mathbf{G}}{dt}\right)_e= \left(\frac{d\mathbf{G}}{dt}\right)_E+\Lambda.\mathbf{G}$$ we get
$$\left(\frac{d\mathbf{G}}{dt}\right)_e=\left(\frac{d\mathbf{G}}{dt}\right)_E+\langle \mathbf{A},\,\mathbf{G}\rangle \mathbf{V}-\langle \mathbf{V},\,\mathbf{G}\rangle \mathbf{A}
+(0,\,\vec{\Omega}_{Th}\wedge\vec{G})$$
Using then $$\left(\frac{d\mathbf{G}}{dt}\right)_e=k\mathbf{V};\qquad\langle \mathbf{A},\,\mathbf{G}\rangle=k;\qquad \langle \mathbf{V},\,\mathbf{G}\rangle=0; $$ we obtain:
$$\left(\frac{d\mathbf{G}}{dt}\right)_E+(0,\,\vec{\Omega}_{Th}\wedge\vec{G})=0$$ 
The left hand side of this equation is the Fermi-Walker derivative of $G$ in the $\mathbf{V}$ direction.
\begin{figure}[h]
\centering
\includegraphics[scale=0.6]{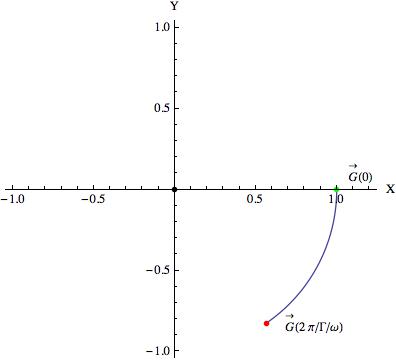}
\caption{\scriptsize{Gyroscope rotation in the plane $(M,E_1,E_2)$ when $M$ goes a complete rotation (360-degree). We used here for $R$ and $\omega$ the values $R=1,\omega=\frac{5}{10}$}}.
\end{figure}
Using
$\left(\frac{d\mathbf{G}}{dt}\right)_E=\frac{d\mathbf{G}_E}{d\tau}=\left(0,\,\frac{d\vec{G}}{d\tau}\right)$
this last equation becomes
\begin{equation}
\label{gyro2}
\frac{d\vec{G}}{d\tau}=-\,\vec{\Omega}_{Th}\wedge\vec{G}
\end{equation}
Consequently, the gyroscope rotates  \textit{with respect to $(M, E)$} in the opposite direction to the instantaneous Thomas rotation. $(M, E)$ taking again its initial orientation after a complete period, this gap shows that the gyroscopes also rotate with respect to the inertial referential frame $(O,\,e$)\\
It can be noted that the solution of (\ref{gyro2}) (with the initial condition $\mathbf{G}_0=(0,\,\vec{G_0})$ also is the Fermi-Walker parallel transport of $\vec{G}_0$ along $\mathcal{L}$  \cite{Straumann} \cite{Ryder}.\\
\begin{figure}[h]
\centering
\includegraphics[scale=0.5]{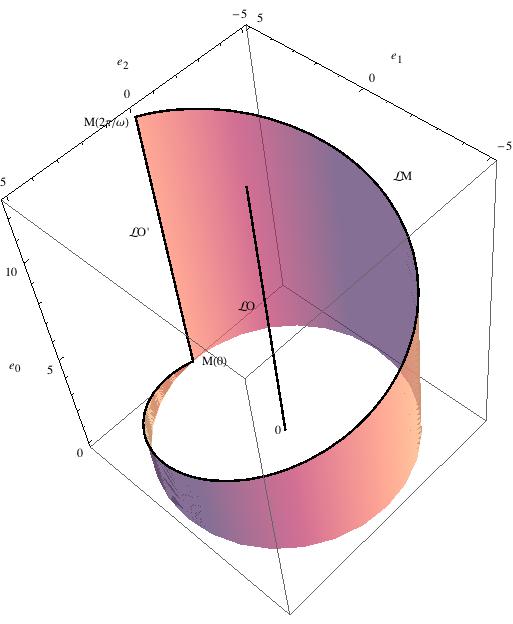}
\caption{\scriptsize{Illustration of "Langevin's twins" in the case of an electron rotating on a circular orbit around the atom nucleus. The twins are denoted $O'$ and $M$ respectively. The straight line parallel to the axis of the cylinder is the worldline of $O'$; the helix is that of $M$.}}
\end{figure}
\section{Conclusion: Langevin's Twins and Thomas precession}
The main results of every dynamical system are contained in the tangent boost $L$ (which gives its 4-velocity $\mathbf{V}$ and the basis vectors of its rest frame $(M,E)$), and its associated matrix of the Lie algebra $\Lambda$ which gives its acceleration and the instantaneous Thomas rotation.\\
The age of the electron with respect to the atom nucleus is then obtained by integrating $L_{1,1}$ over one period $T$. In the case of a uniform circular motion its value is $\Gamma T$\\
The gyroscope rotation $\Psi$ in the physical space $(M, E_1,E_2)$ can be obtained by integrating over one period between $t = \Gamma \tau =0$ and $t = \Gamma \tau =2 \pi/\omega$. We get
\begin{equation*}
\Psi= -2 \pi (\Gamma-1)
\end{equation*}
We thus see that in the case of Langevin's twins, (here, in the case of a uniform circular motion), when the twin who made a journey into space returns home he is not only younger than the twin who stayed on Earth but he is also disorientated with respect to the terrestrial frame because his gyroscope has turned with respect to earth referential frame!


\begin{thebibliography}{99}
\bibitem{Gourgoulhon} E. Gourgoulhon, \textit{Special Relativity in General Frames ; From Particles to Astrophysics}, Springer (Berlin) (2013) or \textit{Relativité restreinte: des particules à l'astrophysique}
EDP Sciences, CNRS Editions (2010).
\bibitem{Parizet} J. Parizet, La géométrie de la relativité restreinte, Ellipses, France, (2008).
\bibitem{Misner} W. Misner S.Thorne J.A. Gravitation, p. 1117, Freeman.
\bibitem{Straumann} N. Straumann, General Relativity with Applications to Astrophysics, p. 50-54, Springer (2004).
\bibitem{Ryder} L. Ryder, Introduction to General Relativity, p. 207-208, Cambridge (2009).
\end{thebibliography}
\end{document}